# Observation of Superoscillation Superlattices

Xin Ma,[1][†] Hao Zhang,[1][†] Wenjun Wei,[1] Yuping Tai,[1] Xinzhong Li,[1,2,3][*] and Yijie Shen[4,5][*]

[1] *School of Chemistry and Chemical Engineering & School of Physics and Engineering, Henan University of Science and Technology, Luoyang 471023, China*
[2] *State Key Laboratory of Transient Optics and Photonics, Xi'an Institute of Optics and Precision Mechanics, Chinese Academy of Sciences, Xi'an, 710119, China*
[3] *Provincial and Ministerial Co-construction of Collaborative Innovation Center for Non-ferrous Metal New Materials and Advanced Processing Technology, Luoyang, 471023, China*
[4] *Centre for Disruptive Photonic Technologies, School of Physical and Mathematical Sciences & The Photonics Institute, Nanyang Technological University, Singapore 637371, Singapore*
[5] *School of Electrical and Electronic Engineering, Nanyang Technological University, Singapore 639798, Singapore*

*Corresponding author: xzli@haust.edu.cn and yijie.shen@ntu.edu.sg

[†]These authors contributed equally to this work.

Superoscillation (SO) wavefunctions, that locally oscillate much faster than its fastest Fourier component, in light waves have enhanced optical technologies beyond diffraction limits, but never been controlled into 2D periodic lattices. Here, we report the 2D superoscillation lattices (SOL) with controlled symmetries, where the local wavevector can be ~700 times larger than the global maximal wavevector ($k_0$) in a localized region ~100 times smaller than the global minimal wavelength ($\lambda_0$). We also demonstrate the superoscillation superlattices (SOSL) as twisted bilayer Moiré patterns of two SOL, akin to the magic angle tuning in advanced twistronics, we can continually tune the on-demand SO with local maximal wavevector in a range of $450k_0$–$700k_0$ and with ~ $\lambda_0/100$–$\lambda_0/1000$. The twistronic SOSL will advance optical imaging and metrology into extreme higher-dimensional superresolution.

In any optical system, the ultimate resolution limitations of light waves subject to the highest frequency in its Fourier decomposition, however, this wisdom has recently been defeated by the phenomenon of *superoscillation* (SO) [1]. A SO wavefunction has the following basic features [1-4]:
(1) It should be band-limited, i.e. with a highest frequency or wavevector in its global spectrum;
(2) It includes a local segment oscillating faster than its highest frequency, i.e., has a local wavevector higher than the maximal global wavevector;
(3) The size of its SO segment, also called hotspot, is smaller than its minimal wavelength component.

Feature (1) is a prerequisite for the manifestation of the SO phenomenon. Features (2) and (3) serve as two metrics to quantify the rate of oscillation whether it is rapid enough beyond the fast Fourier component, evaluated by instantaneous frequency and segment wavelength references, respectively. Consequently, feature (3) is a more strict criterion than (2), and the phenomenon of SO can be generally or specially defined by criteria outlined in (1)&(2) or (1)&(3), respectively. The initial observation of optical SO was reported in focused light fields of nanohole arrays, where many very small optical SO hotspots beyond diffraction limits were quasiperiodically distributed [5,6]. Soon after, by using advanced methods of metamaterials [7-10] and spatial light modulation [11-14], more on-demand SO optical fields can be designed, such as radial spatial mode with controlled hotspot and 1D temporal signal [15-17]. Such 1D or radial SO modes have recently achieved advanced nanometric even picometric optical metrology [1,18,19], superresolution label-free imaging and microscopy [1,20,21], super-sensitivity spectroscopy [17], and so on. However, the manipulation of SO in 2D, especially that of periodic structure, is elusive. There were prior studies of multiple SO hotspots in complex wave interference induced quasiperiodical or speckle patterns [1,5,6,22], but which cannot be observed and controlled in on-demand regular periodic lattice structures, thus prevents the application extension towards higher dimensions.

In this letter, we report the 2D superoscillation lattices (SOL) in light fields. Leveraging a composite system comprising a spatial light modulator (SLM) and an interferometer, we are enabled to conduct precise measurements of the phase and local wavevector distribution inherent to light. This methodology facilitates the detection of SO phenomena in long-range periodic optical lattices and affords the capability to digitally manipulate lattice symmetries. Moreover, inspired by recent advances of twistronics and Moiré superlattices from electronics to photonics [23-28], we introduce the concept of superoscillation superlattices (SOSL) which is achieved by generating Moiré patterns through the superposition of two twisted layers of SOL. Within the context of the twisted bilayer SOSL, we demonstrate the potential for continuous tuning and enhancement of the superoscillation features, which are characterized by their extreme on-demand characteristics, facilitated by the adjustable twisted angle.

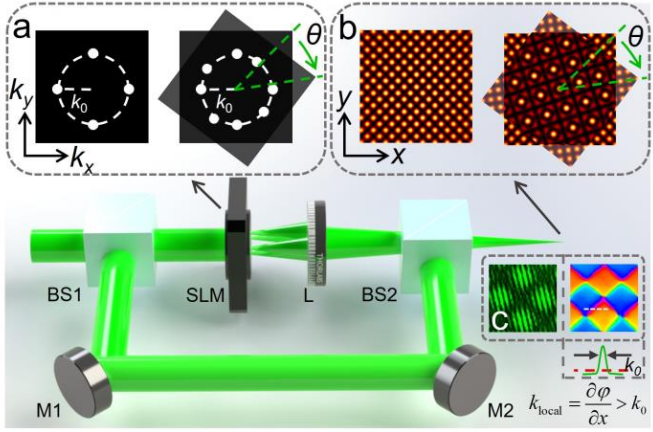

FIG. 1. Production principle and experimental setup: (a) The structures of monolayer and twisted bilayer spatial spectral plane. (b) The corresponding optical lattice intensity patterns generated by the plane-wave-illuminated spectral plane with lens focusing. (c) Experimental interferogram between the SOL and the plane wave, retrieved phase, and $k_{local}$. BS, beam splitter; SLM, spatial light modulator; L, lens; M, mirror.

Initially, we elucidate the underlying principle and the experimental configuration employed for the generation of SOL, as delineated in Fig. 1. A monochromatic plane wave, characterized by a wavelength of 532 nm, is bifurcated into two beams of equivalent energy via beam splitter BS1. Subsequently, one of the beams is directed onto a SLM, which facilitates the digital loading of both monolayer and twisted bilayer spectra, as depicted in Fig. 1(a). In this context, the monolayer spectral plane is composed of a series of distinct dark holes arranged in a periodic array; for illustrative purposes, the number of dark holes is exemplified as $N = 4$, as shown in Fig. 1(a). It is noteworthy that SOL with a regular periodic structure is attainable solely when the count of dark holes is $N = 3, 4$, or $6$, attributable to the inherent symmetry of the dark holes. In contrast, other numerical values result in the formation of quasicrystalline structures, which exhibit angular symmetry but lack a two-dimensional periodic arrangement, as detailed in reference [29]. The bilayer spectral plane was generated by stacking two monolayer spectral plane with controllable twisted angle $\theta$ between the two spectral planes. The radius of the hole determines the maximum frequency component (fastest Fourier component) $k_0 = \max \sqrt{\left(k_x^2 + k_y^2\right)}$, where the size of the hole is relevant small and negligible. So, this satisfies the band-limit feature (1) in introduction for SO: it should be band-limited with a highest frequency or wavevector in its global spectrum. Then, to generate the SOSL, the modulated beam by the SLM pass through a lens to perform the Fourier transform. The intensities of the optical lattices of the SOL and SOSL are recorded at another Fourier spectral plane (focal plane of the lens), respectively, as shown in Fig. 1(b).

To ascertain and validate the SO phenomenon within the optical lattices, the experimental phase was meticulously measured through a precise interferometric process involving the optical lattices and the plane wave (further details are provided in the Supplementary section A). Following the reflection from the two mirrors (M1 and M2) and the BS2, the plane wave reflecting from BS1 engages in interference with the optical lattices that have traversed BS2. The resultant experimental interference intensity and the derived phase patterns are graphically represented in Fig. 1(c). Furthermore, the curve depicted in Fig. 1(c) illustrates the variation in the experimental phase gradient, that is, the local wavevector. The position at which the phase undergoes a transformation is demarcated by a white dashed line in Fig. 1(c). From the curve, the local wave vector at the location (the green curve below the phase pattern) of the white dotted line is greater than the global wave vector (the red dashed line below the phase pattern) $k_{local} = \partial\varphi/\partial x > k_0$. So, this satisfies the feature (2) in introduction for SO: it includes a local segment oscillating faster than the highest frequency, in other words, has a local wavevector higher than the maximal global wavevector. Therefore, the SOL satisfies the general definition of the phenomenon of SO, fulfilling features (1)&(2).

In the subsequent sections, we present the experimental results pertaining to the monolayer SOL with $N = 3, 4, 6$, as illustrated in Fig. 2. The quantity of dark holes, denoted by $N$, on the SLM dictates the symmetry of the SOL. Our research concentrated on two-dimensional regular lattices, yielding structures with triangular, square, and hexagonal sublattices. Supplementary findings for quasicrystalline configurations with $N = 5, 7, 8, 9, 10$ are detailed in Supplementary section B. To substantiate the experimental findings, the simulated intensities are portrayed in Figs. 2(a1)–(c1), where the green dashed lines in each subfigure delineate the three distinct protocell configurations of the optical SOL for $N = 3, 4, 6$, respectively. Subsequently, the complete SOL pattern is constructed by replicating the protocell through translation. Here, the size of the hotspots is measured as $0.93\lambda_0$, $1.43\lambda_0$, $0.26\lambda_0$, respectively, where the wavelength $\lambda_0 = 2\pi/k_0$. Note that the size of the hotspot is smaller the $0.5\lambda_0$, which satisfies the feature (3) for SO: the size of the SO segment, also called hotspot, is smaller than the minimal global wavelength. This validation confirms that the SOL meet the stringent special criteria for the SO phenomenon, fulfilling all features (1–3). Additionally, the corresponding simulated phase patterns are displayed in Fig. 2(a2)–(c2), with the subfigures presenting an enlarged view of the phase within the delineated black dashed box. The green dashed line indicates the rate of phase alteration at the position marked by the black dashed line. Figures 2(a3)–(c3) depict the phase gradient distributions, where the green curves at the base of each figure represent the variation in phase gradient, with the corresponding positions highlighted by green dashed lines. The global maximum wave vector component is indicated by the red dashed line, which is notably smaller in comparison to the local wave vector (green broken line). It is observable that regions of rapid phase gradient change are situated in proximity to the hotspots, and the local wavevector $k_{local} = \partial\varphi/\partial x$ is larger than the amplitude of the wavevector in free space $k_0 = 2\pi/\lambda_0$. The subfigures in Fig. 2(a3)–(c3) show how fast each local wave vector changes relative to its wavelength. Here, the calculation expression is $F = FWHM/\lambda_0$ and FWHM is the value of full width half maximum. The local wave vector at the location is greater than the global wave vector, which satisfies the feature (2) for SO.

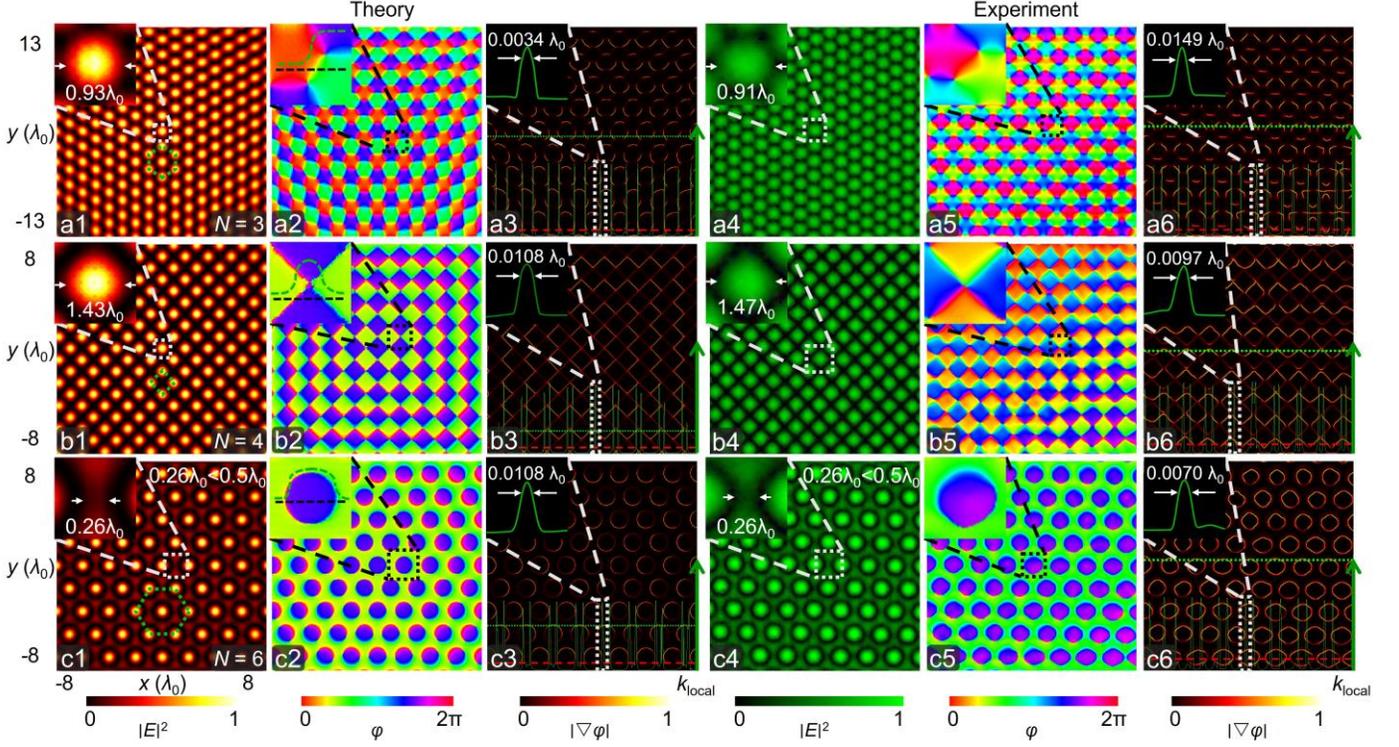

FIG. 2. SOLs with different symmetries ($N = 3, 4$ and $6$): (a1–c1) Intensity patterns, the green dotted lines mark the protocell structure, the insert marks size of SO hotspot as zoom-in of the white dotted line region. (a2–c2) Phase distributions, the insert is zoom-in of the black dotted line region, the green dashed line in the zoom-in insert is the phase change at the position of the black dashed line. (a3–c3) Local wavevector distribution, the red dashed line marks $k_0$, green arrow is axis coordinate local wavevector value, the phase gradient is measured on the green dotted line, the green line is the phase gradient value, the insert marks size of phase gradient value as zoom-in of the white dotted line region, between the two white arrows is FWHM. (a4–c4) Experimental result counterparts of (a1–c1). (a5–c5) Experimental result counterparts of (a2–c2). (a6–c6) Experimental result counterparts of (a3–c3).

The experimental results are presented in the right-hand portion of Fig. 2, encompassing the intensities as depicted in Fig. 2(a4)–(c4), the phases as shown in Fig. 2(a5)–(c5), and the phase gradients as illustrated in Fig. 2(a6)–(c6). These experimental data are found to be in close concordance with the corresponding simulation results. Furthermore, the key characteristics are faithfully reproduced in the experimental results, such as rapid oscillations and nuanced amplitude variations. Among them, the size of the experimental hotspots is $0.91\lambda_0$, $1.47\lambda_0$, $0.26\lambda_0$, respectively. However, the size of hotspot at $N = 4$ is slightly larger than the simulation results owing to the experimental error. The experimental size of the hotspot is smaller the corresponding half of wavelength, which satisfies the feature (3) for SO. The recovered phases were obtained according to the experimental interferogram, as shown in Fig. 3(a5)–(c5). The subfigures in Fig. 2(a6–c6) depicts the experimental phase change, where the local wave vector in the experiment is greater than the global wave vector, and satisfies the feature (2). Here, the red and green dotted lines are $k_0$, and phase gradient change value, respectively. Moreover, the green dotted line in the inserts shows the FWHM. Therefore, the necessary conditions for the experimental generation of SO are satisfied and it is the first time to experimentally generate the 2D SOL.

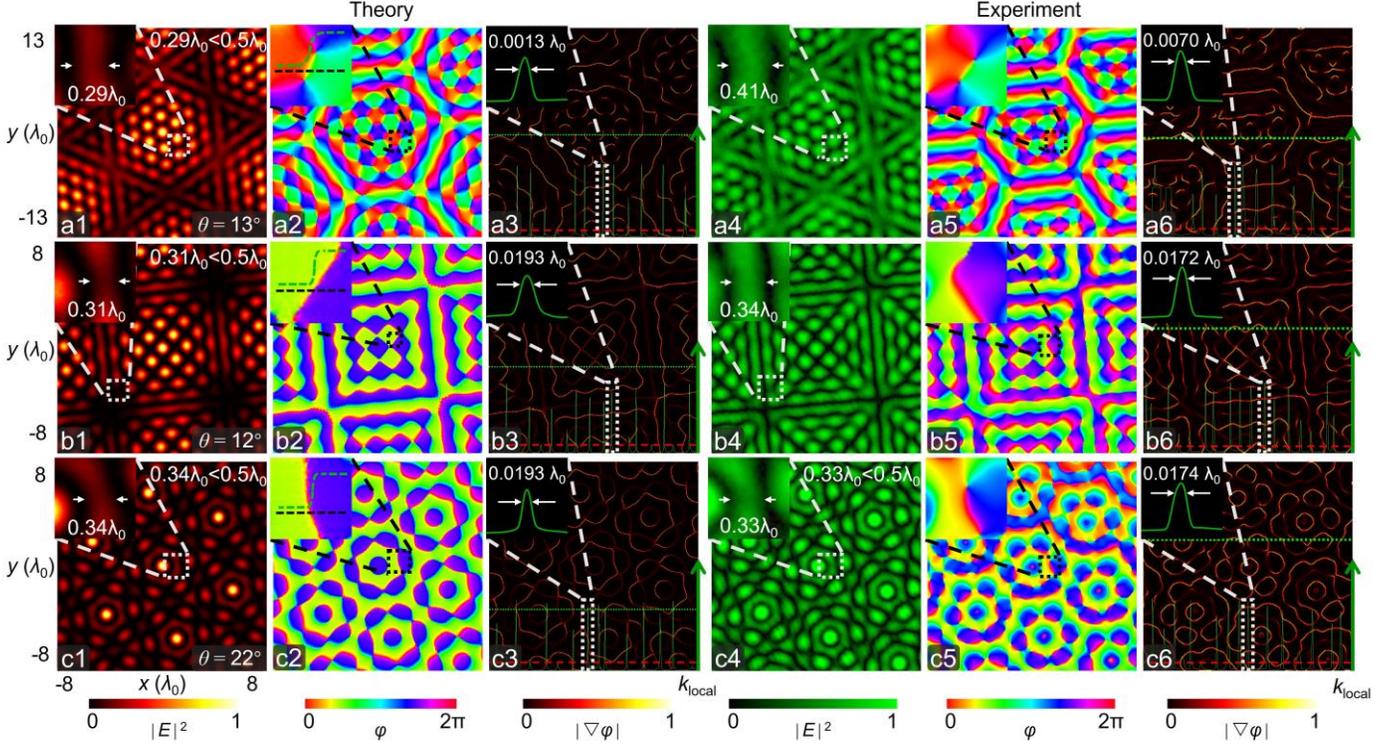

FIG. 3. SOSL with specific twist angle $\theta = 13°$, $12°$ and $22°$ and number of dark holes $N = 3$, 4 and 6. (a1–c1) Intensity patterns, the insert marks size of SOSL hotspot as zoom-in of the white dotted line region. (a2–c2) Simulated phase, the insert is zoom-in of the black dotted line region, the green dashed line in the zoom-in insert is the phase change at the position of the black dashed line. (a3–c3) Local wavevector distribution, the red dashed line marks $k_0$, green arrow is axis coordinate local wavevector value, the phase gradient is measured on the green dotted line, the green line is the phase gradient value, the insert marks size of phase gradient value as zoom-in of the white dotted line region, between the two white arrows is FWHM. (a4–c4) Experimental intensity, the insert is size of hotspot. (a5–c5) Experimental retrieved phase, the insert is the phase change. (a6–c6) Experimental phase gradient. Detailed dynamic modulation process (see Multimedia 1-3).

Beyond the SOL generated by the monolayer spectral plane, drawing an analogy to the distinctive characteristics of twisted graphene, SOSL can be synthesized by introducing an additional degree of freedom, namely the twist angle, to the bilayer spectral plane. The simulated results pertaining to the SOSL are portrayed in Fig. 3, with the twist angles meticulously chosen as $\theta = 13°$, $12°$, and $22°$ for the dark hole counts $N = 3$, 4, 6, respectively. The corresponding intensities are shown in the Figs. 3(a1–c1) and more details about the effect of the twist angle can be seen in Supplementary section C. The size of the simulated hotspots is $0.29\lambda_0$, $0.31\lambda_0$, $0.34\lambda_0$, respectively, which are all smaller than $0.5\lambda_0$ and satisfies the feature (3) for SO. After rotation, the size of the hotspot (inserts in Figs. 3(a1–c1)) remains basically unchanged compared with the results before the rotation. In addition, compared with Fig. 2(a2) and (a3), the phase changes faster and the change of the $F$ is smaller, as shown in Fig. 3(a2) and (a3), respectively. These phenomena indicates that the local wave vector is larger than the global wave vector and satisfies the feature (2) for SO. Moreover, the alterations observed in the SOSL are more pronounced compared to those in the SOL at $\theta = 0°$. Conversely, the phase gradient variations depicted in Fig. 3(b3) and (c3) exceed those observed in Fig. 2(b3) and (c3). Hence, the twist angle introduces supplementary degrees of freedom to the SO phenomenon, allowing for the modulation of the SO range by adjusting the twist angle. The experimental results are displayed in the right-hand section of Fig. 3, with the dimensions of the experimental hotspots measured as $0.41\lambda_0$, $0.34\lambda_0$, and $0.33\lambda_0$, respectively, which are in close concurrence with the simulated outcomes. The experimental phase alterations and phase gradients presented in Fig. 3(a5)-(c5) and Fig. 3(a6)-(c6) are found to be analogous to the simulated results in Fig. 3(a2)-(c2) and Fig. 3(a3)-(c3), respectively. We have successfully achieved, for the first time, the experimental generation of two-dimensional SOSL with controlled oscillation ranges.

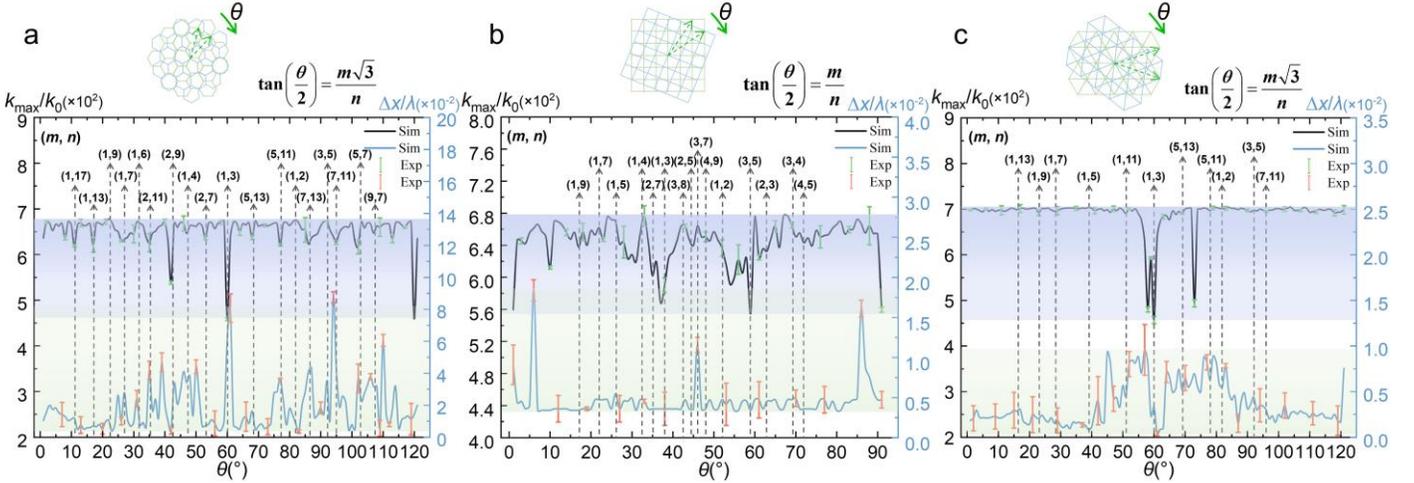

FIG. 4. Two-dimensional Moiré lattice. (a) Six-fold rotation symmetry. (b) Four-fold rotation symmetry. (c) Three-fold rotation symmetry. Top row, schematic discrete representation of two rotated sublattices. Bottom row, analysis of SO characteristics. ($m$, $n$): the minimalist ratio, black and blue curves depict the biggest Fourier components and phase change speed. The shaded areas represent adjustable ranges, green and orange error bars represent the corresponding experimental error range of the two components.

Optical lattice with specific structure which generated by rotating two layers is also called the Moiré lattice. To explore the experimental characteristics of the Moiré lattice, the structure of the optical lattice with different twist angle and fixed input beam was recorded. The results of the Moiré lattice generated by three, four and hexagonal lattices are shown in Fig. 4(a)–(c), respectively. The top row shows the schematic of two rotated sublattices with discrete representation and the SO range is adjustable by rotating the bilayer lattice. Here, the stability can be calculated by the formula [$\tan(\theta/2) = 3^{1/2}m/n$]. The bottom row with some curves was used to analysis the SO characteristics. Black and blue curves depict the biggest Fourier components and the phase change speed, respectively. Corresponding to the crystal lattice, the twist angles that produce the simplest ratio can be given by $\tan(\theta/2) = 3^{1/2}m/n$, $\tan(\theta/2) = m/n$, $\tan(\theta/2) = 3^{1/2}m/n$. Here, $m$ and $n$ are the hypotenuse and height of the triangle formed by the coincidence and center points before and after the rotation, respectively. More details about the effect of the twist angle can be seen in Supplementary section D. This represents the inaugural instance wherein the optical lattice structure has been computationally determined through a formulaic approach. By calculating the values of $m$ and $n$, the Moiré lattice can be generated at specific angles, with the numerical values enclosed in parentheses as depicted in Fig. 4. If the resultant rational number possesses a stable periodicity, the irrational number, conversely, will exhibit an unstable periodicity. This phenomenon, previously identified in reference [29], has not been thoroughly elucidated in prior studies. The shaded regions in Fig. 4 delineate the adjustable ranges of SO, with the three-fold rotation symmetry exhibiting the most extensive SO range, as illustrated in Fig. 4(c). Furthermore, the SO phenomenon becomes observable at special angles, such as at the locations of crests or troughs. The green and orange error bars in the figure signify the corresponding experimental discrepancies, which oscillate within a narrow margin. Above all, we can find 2D SOSL with controlled symmetries, where the local wavevector can be ~700 times larger than the global maximal wavevector, and in a localized region, it can be ~100 times smaller than the global minimal wavelength.

In this work, we generate a new 2D SOL light field, which can promote the development of SO. Moreover, the SO of vector beams provides additional degrees of freedom for SO. And the generation of three layers and multilayer Moiré lattices and other numbers are quasicrystal structure are even more interesting in SO light field [30-33].

In conclusion, we have introduced and successfully generated 2D SOSL through the utilization of monolayer and bilayer spectral planes, which embody the SO phenomenon. Notably, the local oscillation frequency of these lattices significantly surpasses that of their fastest Fourier component. Furthermore, drawing an analogy to twisted graphene, the superposition and torsion of two SOL layers can yield a bilayer Moiré lattice. The attributes of the SO can be modulated, either enhanced or diminished, by adjusting the twist angle. Such SOSL are poised to propel the evolution of superresolution and higher-dimensional capabilities in the realms of optical imaging and metrology. Our SOSL produce the subwavelength hotspot which is smaller than the Abbe's diffraction limit, and can be used for trapping atoms in optical tweezers [34,35].

**Acknowledgments**
This work was supported by the Natural Science Foundation of Henan Province (232300421019, 222300420042); National Natural Science Foundation of China (12274116, 12174089, 12404343); Key Scientific Research Project of Colleges and Universities in Henan Province (21zx002); State Key Laboratory of Transient Optics and Photonics (SKLST202216). Y. Shen acknowledges the support from Nanyang Technological University Start Up Grant and Singapore Ministry of Education (MOE) AcRF Tier 1 grant (RG157/23).

**Reference**

1. N.I. Zheludev, G. Yuan, Optical superoscillation technologies beyond the diffraction limit. Nat. Rev. Phys. **4**, 16–32 (2022).
2. J. Lindberg, Mathematical concepts of optical superresolution. J. Opt. **14**, 083001 (2012).
3. P. Ferreira, A. Kempf, Superoscillations: Faster than the Nyquist rate. IEEE T. Signal Proces. **54**, 3732–3740 (2006).
4. M. Berry, N. Zheludev, Y. Aharonov, et al. Roadmap on superoscillations. J. Opt. **21**, 053002 (2019).
5. F. M. Huang, N. Zheludev, Y. Chen, Javier Garcia de Abajo, F. Focusing of light by a nanohole array. Appl. Phys. Lett. **90**, 091119 (2007).
6. F. M. Huang, Y. Chen, F. J. G. De Abajo, et al. Optical super-resolution through super-oscillations. J. Opt. A-Pure Appl. Op. **9**, S285 (2007).
7. E. T. Rogers, J. Lindberg, T. Roy, et al. A super-oscillatory lens optical microscope for subwavelength imaging. Nat. Mater. **11**, 432-435 (2012).
8. G. H. Yuan, E. T. Rogers, N. I. Zheludev, Achromatic super-oscillatory lenses with sub-wavelength focusing. Light Sci. Appl. **6**, e17036 (2017).
9. G. Chen, Z.Q. Wen, C.W. Qiu, Superoscillation: from physics to optical applications. Light Sci. Appl. **8**, 56 (2019).
10. G. H. Yuan, K. S. Rogers, E. T. F. Rogers, et al. Far-field superoscillatory metamaterial superlens. Phys. Rev. Appl. **11**, 064016 (2019).
11. A. Forbes, A. Dudley, M. McLaren, Creation and detection of optical modes with spatial light modulators. Adv. Opt. Photonics **8**, 200-227 (2016).
12. J. Wu, Z. Wu, Y. He, et al. Creating a nondiffracting beam with sub-diffraction size by a phase spatial light modulator. Opt. Express **25**, 6274-6282 (2017).
13. Z. Wan, Z. Wang, X. Yang, et al. Digitally tailoring arbitrary structured light of generalized ray-wave duality. Opt. Express **28**, 31043-31056 (2020).
14. Y. Shen, Rays, waves, SU (2) symmetry and geometry: toolkits for structured light. J. Opt. **23**, 124004 (2021).
15. E. T. F. Rogers, S. Savo, J. Lindberg, et al. Super-oscillatory optical needle. Appl. Phys. Lett. **102**, 031108 (2013).
16. J. S. Diao, W. Z. Yuan, Y. T. Yu, et al. Controllable design of super-oscillatory planar lenses for sub-diffraction-limit optical needles. Opt. Express **24**, 1924–1933 (2016)
17. G. McCaul, P. Peng, M. O. Martinez, et al. Superoscillations deliver superspectroscopy. Phys. Rev. Lett. **131**, 153803 (2023).
18. G. H. Yuan, N. I. Zheludev, Detecting nanometric displacements with optical ruler metrology. Science **364**, 771-775 (2019).
19. T. Liu, C. H. Chi, J. Y. Ou, et al. Picophotonic localization metrology beyond thermal fluctuations. Nat. Mater. **22**, 844-847 (2023).
20. F. Qin, K. Huang, J. F. Wu, et al. A supercritical lens optical label-free microscopy: sub-diffraction resolution and ultra-long working distance. Adv. Mater. **29**, 1602721 (2017).
21. E. T. F. Rogers, S. Quraishe, K. S. Rogers, et al. Far-field unlabeled super-resolution imaging with superoscillatory illumination. APL Photonics **5**, 066107 (2020).
22. M. R. Dennis, A. C. Hamilton, J. Courtial, Superoscillation in speckle patterns. Opt. Lett. **33**, 2976-2978 (2008).
23. Y. Cao, V. Fatemi, S. Fang, et al. Unconventional superconductivity in magic-angle graphene superlattices. Nature **556**, 43-50 (2018).
24. J. M. Park, Y. Cao, K. Watanabe, et al. Tunable strongly coupled superconductivity in magic-angle twisted trilayer graphene. Nature **590**, 249-255 (2021).
25. P. Wang, Y. Zheng, X. Chen, et al. Localization and delocalization of light in photonic moiré lattices. Nature **577**, 42–46 (2020).
26. G. Hu, Q. Ou, G. Si, et al. Topological polaritons and photonic magic angles in twisted α-MoO$_3$ bilayers. Nature **582**, 209–213 (2020).
27. Q. Zhang, G. Hu, W. Ma, et al. Interface nano-optics with van der Waals polaritons. Nature **597**, 187–195 (2021).
28. L. Du, M. R. Molas, Z. Huang, et al. Moiré photonics and optoelectronics. Science **379**, eadg0014 (2023).
29. P. Wang, Q. Fu, V. Konotop, Observation of localization of light in linear photonic quasicrystals with diverse rotational symmetries. Nat. Photonics **18**, 224-229 (2024).
30. H. Kim, Y. Choi, É. Lantagne-Hurtubise, et al. Imaging inter-valley coherent order in magic-angle twisted trilayer graphen. Nature **623**, 942-948 (2023).
31. P. Tuan, L. Huang, Optical vector fields with kaleidoscopic quasicrystal structures by multiple beam interferenc. Opt. Express **31**, 33077-33090 (2023).
32. Z. Ji, Y. Zhao, Y. Chen, et al. Opto-twistronic Hall effect in a three-dimensional spiral lattice. Nature (2024). https://doi.org/10.1038/s41586-024-07949-1
33. L. Du, Z. Huang, J. Zhang, et al. Nonlinear physics of moiré superlattices, Nat. Mater. **23**, 1179-1192 (2024).
34. H.M. Rivy, S.A. Aljunid, E. Lassalle, et al. Single atom in a superoscillatory optical trap. Commun. Phys. **6**, 155 (2023).
35. Z. Meng, L. Wang, W. Han, et al. Atomic Bose–Einstein condensate in twisted-bilayer optical lattices. Nature **615**, 231-236 (2023).